# Epitaxial Growth of Rutile GeO$_2$ via MOCVD

Imteaz Rahaman[1], Bobby Duersch[2], Hunter D. Ellis[1], Michael A. Scarpulla[1,3], and Kai Fu[1, a)]

[1]*Department of Electrical and Computer Engineering, The University of Utah, Salt Lake City, UT 84112, USA*

[2]*Electron Microscopy and Surface Analysis Laboratory, The University of Utah, Salt Lake City, UT 84112, USA*

[3]*Department of Materials Science and Engineering, The University of Utah, Salt Lake City, UT 84112, USA*

**Abstract**

Rutile Germanium Dioxide (r-GeO$_2$) has been identified as an ultrawide bandgap (UWBG) semiconductor recently, featuring a bandgap of 4.68 eV—comparable to Ga$_2$O$_3$—but offering bipolar dopability, higher electron mobility, higher thermal conductivity, and higher Baliga's figure of merit (BFOM). These superior properties position GeO$_2$ as a promising material for various semiconductor applications. However, the epitaxial growth of r-GeO$_2$, particularly in its most advantageous rutile polymorph, is still at an early stage. This work explores the growth of r-GeO$_2$ using metal-organic chemical vapor deposition (MOCVD) on an r-TiO$_2$ (001) substrate, utilizing tetraethyl germane (TEGe) as the precursor. Our investigations reveal that higher growth temperatures significantly enhance crystalline quality, achieving a full width at half maximum (FWHM) of 0.181º at 925 ºC, compared to 0.54º at 840 ºC and amorphous structures at 725 ºC. Additionally, we found that longer growth durations increase surface roughness due to the formation of faceted crystals. Meanwhile, adjusting the susceptor rotation speed from 300 RPM to 170 RPM plays a crucial role in optimizing crystalline quality, effectively reducing surface roughness by approximately 15 times. This study offers a foundational guide for optimizing MOCVD growth conditions of r-GeO$_2$ films, emphasizing the crucial need for precise control over deposition temperature and rotation speed to enhance adatom mobility and effectively minimize the boundary layer thickness.

---

a) Author to whom correspondence should be addressed. Electronic mail: kai.fu@utah.edu

## I. Introduction

Wide bandgap (WBG) and ultra-wide bandgap (UWBG) semiconductors, such as SiC (3.2 eV),[1] GaN (3.39 eV),[2] AlGaN (3.4-6 eV)[3], β-$Ga_2O_3$ (4.5-4.8 eV),[4] and diamond (5.47 eV)[5], have undergone extensive exploration for applications in power electronics, extreme environment electronics, gas sensors, and UV detectors due to their larger breakdown strength, increased power density, and reduced energy loss.[6–8] β-$Ga_2O_3$ stands out in high-power device applications, evident in Baliga's figure of merit comparison with Si (3000×), SiC (10×), and GaN (4×).[9] However, β-$Ga_2O_3$ has difficulty in effective p-type conductivity and poor thermal conductivity.[10–12] For AlGaN, acceptor activation efficiency, and hole mobility are suppressed by increasing the Al content.[13,14] Diamond suffers from a lack of industrially applicable substrates and effective dopants due to high ionization energies.[15] The advancement of all current UWBG semiconductors is hindered by challenges such as limited wafer sizes and difficulties in doping. Recently, r-$GeO_2$ has emerged as a novel UWBG semiconductor due to its remarkable physical properties and potential to address these challenges. It has a bandgap of 4.44 to 4.68 eV[16–18] and a high predicted electron mobility [244 cm²/Vs ($\perp \vec{C}$) and 377 cm²/V·s (($\parallel \vec{C}$)][19], giving it a high n-type Baliga figure of merit (27,000–35,000 ×$10^6$ $V^2$ $Ω^{-1}$ $cm^{-2}$).[20] Moreover, the thermal conductivity of r-$GeO_2$ is 37 and 58 W/m·K along the *a* and *c* directions, respectively, with a measurement of 51 $Wm^{-1}K^{-1}$, roughly twice that of $Ga_2O_3$.[21] Theoretical studies suggest that r-$GeO_2$ can be ambipolar doped and has relatively high hole mobility [27 $cm^2$/Vs ($\perp \vec{C}$)) and 29 ($\parallel \vec{C}$)], indicating its potential for p-n homojunction and other bipolar devices.[17,19,22] Furthermore, bulk r-GeO$_2$ can be synthesized like Ga$_2$O$_3$ and GeO$_2$, making large wafer sizes of GeO$_2$ available soon.[23–25]

In 2020, Chae *et al.* epitaxially stabilized r-GeO$_2$ thin films on R-plane sapphire substrates using a (Sn,Ge)O$_2$ buffer layer via molecular beam epitaxy (MBE).[26] Their work revealed a narrow growth window and a low growth rate of 10 nm/h for r-GeO$_2$. In 2021, Takane *et al.* established a growth process via mist chemical vapor deposition (mist CVD), achieving growth rates of 1.2–1.7 µm/h on (001) r-TiO$_2$ substrates.[27] Deng *et al.* used pulsed laser deposition to synthesize GeO$_2$ films, facing significant challenges in stabilizing the rutile phase due to the presence of metastable glass forms.[28] In 2024, Rahaman *et al*. reported the growth of polycrystalline GeO$_2$ on R-plane and C-plane sapphires by metalorganic chemical vapor deposition (MOCVD).[29] Despite these advances in thin-film r-GeO$_2$ growth, as far as we know, there is no report so far of the successful growth of r-GeO$_2$ by MOCVD which remains the preferred method for large-scale wafer epitaxy in mass production.

In this study, we report the successful growth of r-GeO$_2$ on (001) TiO$_2$ substrates using MOCVD. We have investigated the impact of growth parameters—temperature (725 ºC to 925 ºC), duration (90 to 180 minutes), and rotation speed (170 RPM to 300 RPM)—on the epitaxial film quality. The films were thoroughly analyzed using X-ray Diffraction (XRD), Scanning Electron Microscopy (SEM), X-ray Reflectivity (XRR), Energy Dispersive X-ray Spectroscopy (EDX), Reciprocal Space Mapping (RSM), and Atomic Force Microscopy (AFM). Our results have shown that high-quality film can be obtained at a high temperature of 925 ºC in this work, with a 170 RPM rotation speed, where the full width at half maximum (FWHM) is 0.181º (651.6 arcsec). Moreover, this condition can result in a relatively low surface RMS roughness of 13 nm. This research offers a guide for growing r-GeO$_2$ single crystal films by MOCVD.

## II. Experimental Details

The growth of GeO$_2$ films was carried out in an Agilis MOCVD system, developed by Agnitron Technology. The r-TiO$_2$ was selected as the substrate due to its relatively small lattice mismatch and low strain with GeO$_2$.[27,30] The growth temperature varied from 725°C to 925°C while maintaining a constant chamber pressure of 80 Torr. TEGe and pure oxygen (O$_2$) were used as precursors, with argon (Ar) serving as both the carrier and shroud gas. The oxygen flow rate was set at 2000 SCCM, the argon shroud gas flow rate at 1250 SCCM, and the TEGe precursor flow rate at 160 SCCM. The susceptor rotation speeds tested were 300 RPM and 170 RPM, to assess the effects of it on film quality. Prior to loading into the MOCVD chamber, the TiO$_2$ substrates were cleaned using a piranha solution (H$_2$SO$_4$ = 3:1), followed by successive cleaning with acetone, isopropanol, and deionized water. A comprehensive list of the growth parameters and their corresponding sample IDs is presented in Table I.

**Table I.** Summary of GeO$_2$ films grown at different growth conditions, including growth temperature, chamber pressure, TEGe precursor flow rate, O$_2$ flow rate, shroud gate flow rate, and susceptor rotation speed.

| Sample ID | Temperature (°C) | Pressure (Torr) | TEGe Precursor Flow rate (SCCM) | Oxygen Flow rate (SCCM) | Shroud gas Flow rate (SCCM) | Rotation speed (RPM) | Growth duration (min) |
|---|---|---|---|---|---|---|---|
| T19 | 725 | 80 | 160 | 2000 | 1250 | 300 | 90 |
| T23 | 840 | 80 | 160 | 2000 | 1250 | 300 | 90 |
| T22 | 925 | 80 | 160 | 2000 | 1250 | 300 | 180 |
| T25 | 925 | 80 | 160 | 2000 | 1250 | 300 | 90 |
| T26 | 925 | 80 | 160 | 2000 | 1250 | 170 | 180 |

The structural properties of the GeO$_2$ thin film were analyzed using a Bruker D8 Discover High-resolution XRD (HRXRD) with a 1.5406 Å Cu Kα$_1$ source with a hybrid monochromator. The surface morphology of the films was examined with the FEI TENEO SEM and the Bruker Dimension ICON AFM.

### III. Results and Discussion

The tetragonal crystal structure of the r-GeO$_2$ is shown in Fig 1 (a). It is thermodynamically the most stable phase of GeO$_2$ with the *a* and *c* lattice parameters of 4.40656 Å and 2.86186 Å, respectively, and not water-soluble.[31] Figure 1(b) depicts the breakdown electric field as a function of the energy bandgap in semiconductors ($E_c \sim E_g^{2.3}$), with GeO$_2$ included.

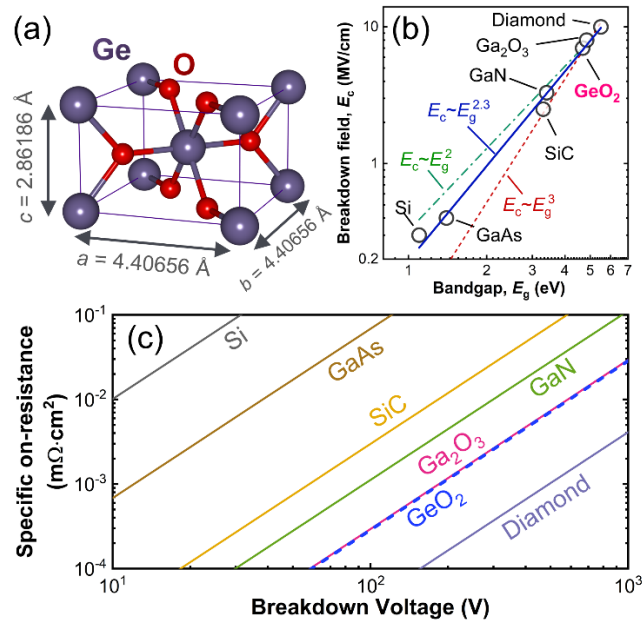

**Fig 1.** (a) Tetragonal crystal structure of r-GeO$_2$, showing the lattice parameters with germanium (Ge) atoms in red and oxygen (O) atoms in purple. (b) Breakdown electric field ($E_c$) vs energy bandgap ($E_g$) for various semiconductor materials. (c) Specific on-resistance of the semiconductors as a function of breakdown voltage based on the BFOM values in Table II.

The plot reveals a clear linear trend of their relationship on a logarithmic scale. Figure 1(c) presents the BFOM of GeO$_2$ alongside other semiconductors, revealing that GeO$_2$ exhibits a slightly higher BFOM than Ga$_2$O$_3$. The calculation of BFOM is highly sensitive to material properties, resulting in variations in the reported values since researchers use different material properties based on their understanding and the quality of the materials continues to advance. The material properties of GeO$_2$ and other semiconductors are listed in Table I based on reported data [32–34] for comparison.

**Table II.** Material properties of GeO$_2$ and other semiconductors.

| Properties | Si | GaAs | 4H-SiC | GaN | GeO$_2$ | Ga$_2$O$_3$ | Diamond |
|---|---|---|---|---|---|---|---|
| Bandgap, $E_g$ (eV) | 1.1 | 1.4 | 3.3 | 3.4 | 4.68 | 4.85 | 5.5 |
| Breakdown field, $E_c$ (MVcm$^{-1}$) | 0.3 | 0.4 | 2.5 | 3.3 | 7 | 8 | 10 |
| Electron mobility, $\mu$ (cm$^2$V$^{-1}$s$^{-1}$) | 1400 | 8000 | 1000 | 1250 | 377 | 300 | 2000 |
| Relative dielectric constant, $\varepsilon$ | 11.8 | 12.9 | 9.7 | 9 | 12.2 | 10 | 5.5 |
| Thermal conductivity, $\lambda$ (Wcm$^{-1}$K$^{-1}$) | 1.5 | 0.5 | 4.9 | 2.3 | 0.51 | 0.1-0.3 | 20 |
| BFOM, $\varepsilon\mu E_c^3$ (GWcm$^{-2}$) | 0.04 | 0.58 | 13.42 | 35.80 | 139.68 | 136.00 | 973.96 |

The crystal quality of the GeO$_2$ films is characterized by HRXRD scanning shown in Fig. 2 (a-c), illustrating the variations in film properties across a range of experimental conditions. It can be seen that the GeO$_2$ film is amorphous for a growth temperature of 725 °C (Fig. 2(a), sample T-19). However, when the growth temperature reaches 840 °C, the film begins to exhibit epitaxial growth of crystalline r-GeO$_2$, indicated by a broad peak (65.14°) in the XRD pattern (Fig. 2(a), sample T-23). This peak sharpens significantly for samples with a growth temperature increasing to 925 °C (Fig. 2(a), sample T-25). This change can be attributed to the enhanced mobility of adatoms at elevated temperatures, which, at lower temperatures, is insufficient to facilitate film crystallization.[26] Additionally, Takane *et al.* have highlighted the critical need to balance adsorption and desorption for successful crystalline film formation.[27] This equilibrium also aids in determining whether the deposited layer will be amorphous or crystalline by comparing the net mass addition rate (growth rate) to adatom diffusivity.

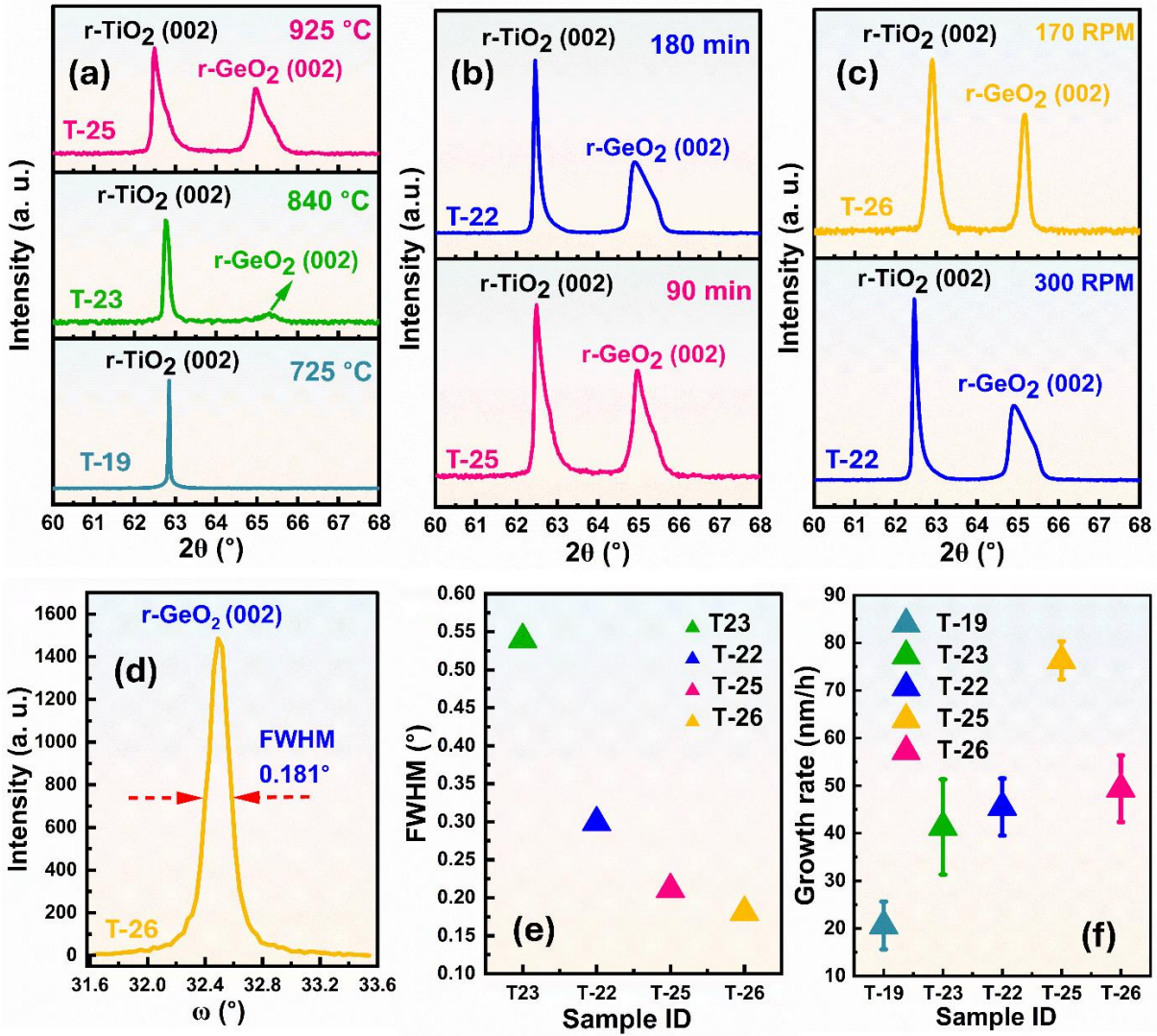

**FIG. 2.** The HRXRD θ-2θ data from epitaxial r-GeO$_2$ Films on (001) r-TiO$_2$ Substrates as a function of (a) growth temperatures with rotation speed of 300 RPM and growth duration 90 min, (b) different growth durations at 925 °C with rotation speed of 300 RPM, and (c) various susceptor rotation speeds at 925 °C with a growth duration of 180 min. (d) XRD rocking curve of the r-GeO$_2$ (002) peak from sample ID T-26. (e) Comparative FWHM of all samples. (f) Growth rates of the samples with different growth conditions. Thickness measurements were conducted by XRR.

For samples with an extended growth period of up to 180 minutes (sample T-22), the XRD peak broadens with an FWHM of 0.299°, compared to 0.211° of sample T-25 with a shorter growth time. Since our previous work has shown that the rotation speed affects the film quality[29], the susceptor rotation speed was also investigated by reducing it from 300 RPM to 170 RPM, maintaining a 180-minute growth period at 925 °C. This adjustment resulted in a sharper XRD peak for r-GeO$_2$ film and an improved FWHM of 0.181° (651.6 arcsec), as shown in the rocking curve for sample T-26 in Fig. 2(d). Figure 2(e) summarizes the FWHM for all the samples. The growth rates of all the samples at different growth conditions are shown in Fig. 2(f), revealing the lowest growth rate for samples with a growth temperature of 725 °C (T-19) and increased growth rates for higher temperatures. The sample T-26, with the best crystalline quality in this study, shows a growth rate of about ~49.36 ± 7 nm/h, which is slightly smaller than the film grown with a higher rotation speed (T-25). The error bar of growth rate was set using Leptos 7.14 software by estimating the film thickness at different angles from the fringe peaks in the Fast Fourier Transform (FFT) analysis.

Figure 3 presents symmetrical and asymmetrical RSMs for the sample T-25 and T-26. The symmetric reflections in Figure 3(a) and (c) align precisely with the corresponding substrate peaks in the $q_\parallel$- direction, indicating minimal tilt of the *c*-axis of the film relative to that of the substrate. The presence of K$\alpha_1$ splitting in the film reflections and minimal broadening in the $q_\perp$- direction in Fig. 3(a) for sample T-26 points to the high quality of the film, in contrast to the observations in Fig. 3(c) for Sample T-25. Furthermore, the higher broadening observed in the $q_\parallel$- direction in Fig. 3(a) suggests slightly increased mosaicity.

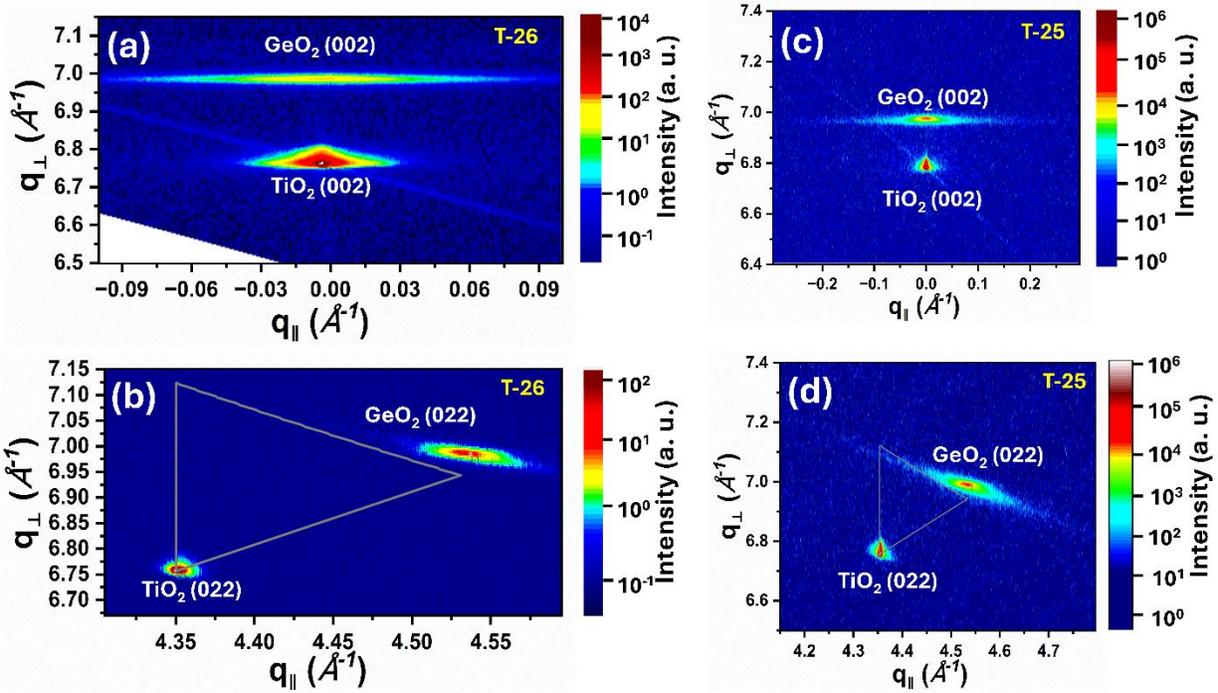

**FIG. 3.** Reciprocal space mappings of TiO$_2$ substrates and r-GeO$_2$ films: (a) symmetrical (002) reflections for sample T-26, (b) asymmetrical (022) reflections for sample T-26, (c) symmetrical (002) reflections for sample T-25, and (d) asymmetrical (022) reflections for sample T-25. Data for these RSM plots were generated and analyzed using Leptos 7.14 software.

Additionally, the (022) reflections of Fig. 3(b) exhibit low broadening with Kα$_1$ splitting in the $q\perp$-direction, indicating less elongation in regions with coherent scattering laterally. Nevertheless, the presence of the (022) reflections in their correct positions relative to the substrate reflections, along with similar $q_\parallel$ values, confirms the epitaxial growth of the thin film. However, despite these similarities, the $q_\parallel$ and $q\perp$ values of these reflections do not align with the corresponding substrate peak positions, as marked by the triangles in Fig. 3(b) and (d). The triangle is automatically generated by the Leptos 7.14 software during our RSM image processing to facilitate the strain analysis of the film. This discrepancy indicates that the (022) reflections of the film are shifted away from the substrate peaks. We have depicted the strain of all the films in Table III.

**Table III.** Characteristics of all the samples estimated from XRD data.

| Sample ID | $2\theta_{r-GeO2}$ (°) | Lattice parameter, $c_{r-GeO2}$ (Å) | Strain of $c$-axis (%) |
|---|---|---|---|
| T-23 | 65.31060 | 2.8552 | -0.23 (compressive) |
| T-22 | 64.90787 | 2.8709 | +0.32 (tensile) |
| T-25 | 64.96467 | 2.8687 | +0.24 (tensile) |
| T-26 | 65.17339 | 2.8605 | -0.048 (compressive) |

The lattice constant $c$ is measured by the following equation [35]-

$$\frac{1}{d_{hkl}^2} = \frac{h^2+k^2}{a^2} + \frac{l^2}{c^2} \tag{2}$$

In this regard, $a$ and $c$ represent the lattice constants, while $d_{hkl}$ refers to the crystalline interplanar distance, which is calculated by Bragg's Law. Specifically, when applying this equation to the (002) diffraction peak, the equation simplifies to $c = 2\, d_{hkl}$. The strain $\varepsilon$, along the $c$-axis is calculated by the equation (3) [36]-

$$\varepsilon = \frac{c-c_0}{c_0} \times 100\% \tag{3}$$

Where $c_0$ (2.86186 Å) is the unstrained $c$ lattice parameter of r-GeO$_2$.[31] Table III shows that the lattice constant $c$ increases with higher growth temperatures, rotation speeds, and growth periods. Specifically, films grown at 840°C (T-23) have a lattice constant of 2.8552Å, which rises to 2.8687Å at 925°C (T-25) under the same conditions of 300 RPM and 90 min. Extending the growth period to 180 min (T-22) further increases the value to 2.8709Å. However, reducing the rotation speed to 170 RPM, sample T-26 exhibits a lattice constant of 2.8605 Å, closely matching the standard 2.86186 Å and showing a compressive strain of 0.048%. The variations in the lattice parameter $c$ can be linked to several factors. At higher temperatures, increased atomic vibrations due to enhanced thermal energy led to thermal expansion and potential alterations in defect densities, both of which can distort the lattice. At a moderate rotation speed of 170 RPM, optimal precursor distribution and sufficient atomic settling time promote uniform film growth. Conversely, too high a rotation speed might decrease the pressure, resulting in a shorter interaction time between precursors and the substrate. Consequently, the moderate rotation speed of 170 RPM produced films with lattice constants near standard values, indicating well-controlled growth conditions that stabilize the crystal structure. Moreover, the growth rate under mass transport-limited conditions is inversely related to the thickness of the boundary layer and varies with the rotation rate and operating pressure.[37] A higher growth rate with a thinner boundary layer at high rotation speed may negatively affect film quality.

Figures 4(a-h) and 5(a-f) depict SEM and AFM analyses of our films under various conditions. Sample T-19, grown at 725 ºC, has a smooth surface as shown in Fig. 4(a). At 840 ºC (T-23), the surface becomes rougher with visible irregular lines (Fig. 4(b)). AFM data show surface roughness of 3.13 nm for T-19 and 6.34 nm for T-23. At 925 ºC, as indicated in Fig. 5(a) and (b). The surface of sample T-22 features distinct faceted crystals (Fig. 4(c)), and prolonged growth times lead to a larger, coalesced square pattern, leading to increased roughness of 181.8 nm (180 min, T-22) compared to ~99.52 nm for shorter periods (90 min, T-25) as shown in Fig. 5(c) and 5(d). In contrast, the surface morphology of sample T-26 with a lower susceptor rotation speed (170 RPM), depicted in Fig. 4(e), shows smaller roughness. The RMS roughness of T-26 is markedly improved to approximately 13 nm, as highlighted in Fig. 5(e). Figure 5(f) compares the RMS roughness of all the samples. The presence of a square pattern in the film raises questions about whether this is due to island growth or defects. To clarify this, EDX analysis was performed on three different regions as shown in Figure 4(f), with the chemical composition of these areas detailed in Fig. 4(g)

and the EDX mapping presented in Fig. 4(h). The comparison of SEM and EDX reveals that both the square-patterned and non-patterned regions exhibit similar ratios of Ge element content. It is worth noting that sample T-23, which lacks the square pattern, still demonstrates the crystalline properties of r-GeO$_2$, albeit of lower quality. These analyses indicate that the square patterns are more likely defects. However, for further confirmation, additional studies such as Transmission Electron Microscopy (TEM) could provide definitive evidence of the epitaxial nature of the film.

Although lower rotation speed here can improve the film quality, excessively low speeds should be avoided as they may lead to amorphous film formation, according to our recent findings.[29] According to classical nucleation theory, surface area scales with the square of the size ($r^2$) and volume scales with the cube ($r^3$) in three-dimensional systems. However, in thin films, both surface area and volume scale are similar to $r^2$. This scaling suggests that interface energy plays a dominant role in shaping the morphology of the film. As a result, this energy influence is also likely responsible for the characteristic faceted or spherulitic patterns observed in different crystalline phases of GeO$_2$ films.

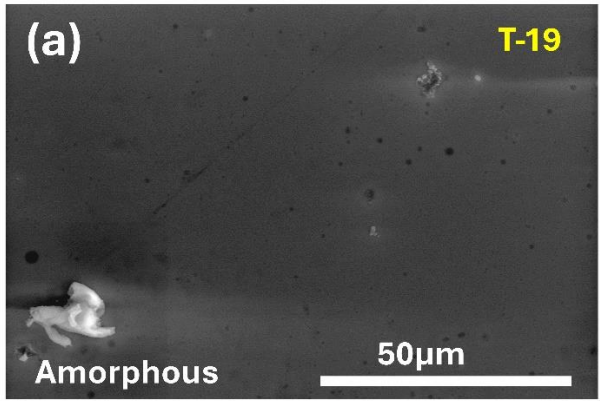
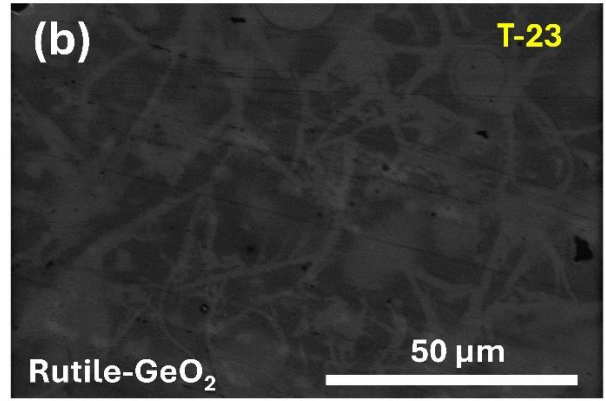
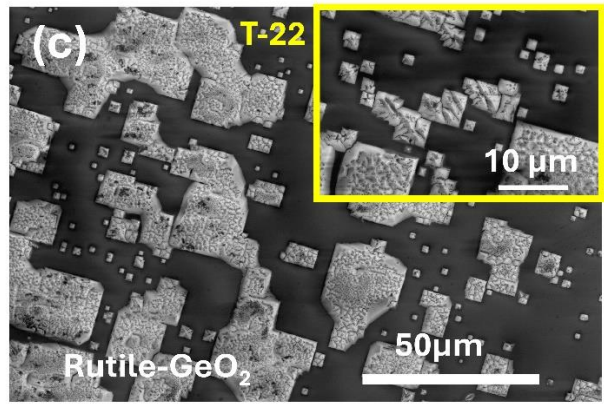
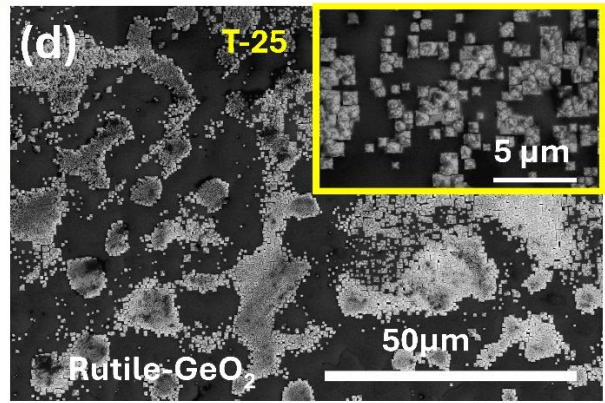
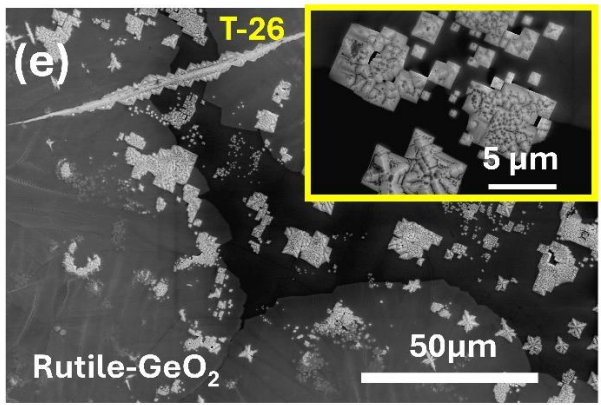
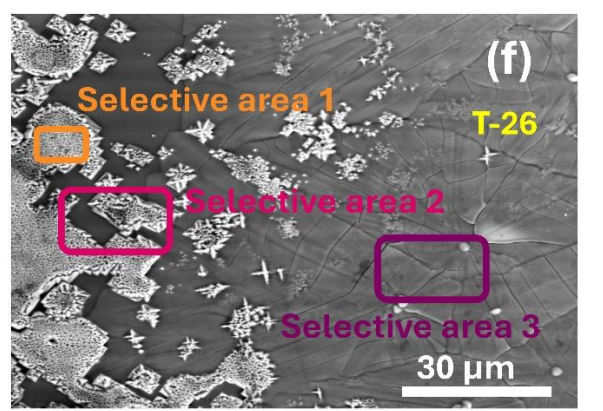
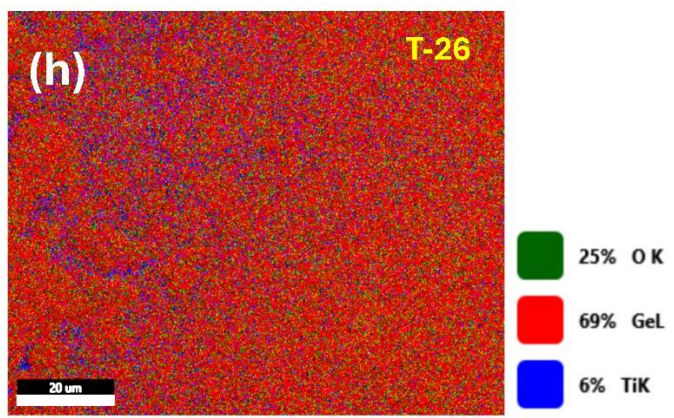

Figure 4. SEM images of r-GeO$_2$ films under various growth conditions: sample ID- (a) T-19, (b) T-23, (c) T-22, (d) T-25, and (e) T-26. Additional analyses for sample T-26 include (f) SEM image for chemical composition analysis, (g) corresponding EDX data for Fig (f), and (h) corresponding EDX mapping for Fig (f).

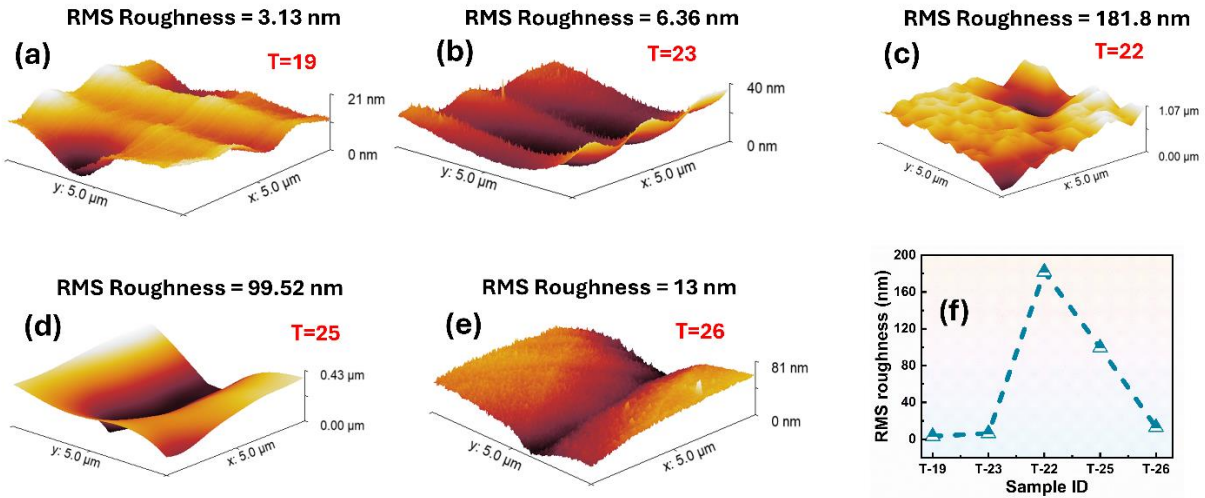

FIG. 5. Surface roughness by AFM of r-GeO$_2$ films at various growth conditions: (a) T-19, (b) T-23, (c) T-22, (d) T-25, and (e) T-26.

### IV. Conclusion

In conclusion, our study has reported the successful growth of single crystal r-GeO$_2$ by MOCVD by investigating different growth parameters. The results affirm that the temperature of growth and rotation speed exert a profound influence on the crystalline quality of GeO$_2$ films. The growth temperature of 925 ºC resulted in high-quality crystal films, while the low temperature of 725 ºC resulted in amorphous based on the XRD results. However, longer growth at 925 ºC makes both the crystal quality and surface morphology worse. By reducing the susceptor rotation speed to 170 RPM, both the epilayer crystal quality and surface roughness were improved, resulting in an FWHM of 0.181º and surface roughness of 13 nm. This comprehensive exploration provides a foundational understanding for achieving a single crystal r-GeO$_2$ epilayer by MOCVD, helping to pave the way for the development of this new UWBG semiconductor.

## AUTHOR DECLARATIONS

### Conflict of Interest

The authors have no conflicts to disclose.

### Author Contributions

Imteaz Rahaman: Data curation (lead); Formal analysis (lead); Investigation (lead); Methodology (lead); Bobby Duersch: Data curation (equal); Writing – review & editing (supporting), Formal analysis (supporting). Hunter D. Ellis: Writing – review & editing (supporting); Michael A. Scarpulla: Conceptualization (equal); Writing – review & editing (equal). Kai-Fu: Conceptualization (lead); Supervision (lead); Project administration (lead); Resources (lead).

## DATA AVAILABILITY

The data that support the findings of this study are available from the corresponding authors upon reasonable request.